\documentclass[traditabstract,bibyear]{aa}
\usepackage{times}
\usepackage{longtable}
\usepackage{natbib}
\usepackage{rotate}
\usepackage{lscape}
\usepackage{amssymb}
\usepackage{graphicx}
\usepackage[draft]{hyperref}

\bibpunct{(}{)}{;}{a}{}{,} 

\def\5{\footnotesize V\normalsize}
\def\4{\footnotesize IV\normalsize}
\def\3{\footnotesize III\normalsize}
\def\2{\footnotesize II\normalsize}
\def\1{\footnotesize I\normalsize}

\def\lam{$\lambda$}
\def\kms{$\mbox{km s}^{-1}$}

\def\o{$\phantom{1}$}

\begin{document}

\title{First stellar spectroscopy in Leo~P}

\author{C.~J.~Evans\inst{1}, N.~Castro\inst{2,3}, O.~A.~Gonzalez\inst{1}, M.~Garcia\inst{4},
N.~Bastian\inst{5}, M.-R.~L.~Cioni\inst{3}, J.~S.~Clark\inst{6},\\ 
B.~Davies\inst{5}, A.~M.~N.~Ferguson\inst{7}, S.~Kamann\inst{5}, D.~J.~Lennon\inst{8,9}, 
L.~R.~Patrick\inst{9,10}, J.~S.~Vink\inst{11}, D.~R.~Weisz\inst{12}}
\offprints{chris.evans@stfc.ac.uk}
\authorrunning{C.~J.~Evans et al.}
\titlerunning{MUSE observations of Leo~P}

\institute{UK Astronomy Technology Centre, Royal Observatory, Blackford Hill, Edinburgh, EH9 3HJ,~UK
  \and
Department of Astronomy, University of Michigan, 1085 S. University Avenue, Ann Arbor, MI 48109-1107, USA 
  \and
Leibniz-Institut f\"{u}r Astrophysik Potsdam, An der Sternwarte 16, D-14482 Potsdam, Germany
  \and
Centro de Astrobiolog\'{i}a, CSIC-INTA, Ctra. Torrej\'{o}n a Ajalvir km.4, 
E-28850 Torrej\'{o}n de Ardoz, Madrid, Spain
  \and
Astrophysics Research Institute, Liverpool John Moores University, Liverpool L3 5RF, UK
  \and
Department of Physics and Astronomy, The Open University, Walton Hall, Milton Keynes, MK7 6AA, UK
  \and
Institute for Astronomy, University of Edinburgh, Blackford Hill, Edinburgh EH9 3HJ, UK
  \and
ESA, European Space Astronomy Centre, Apdo. de Correos 78, E-28691 Villanueva de la Ca\~{n}ada, 
Madrid, Spain
  \and
Instituto de Astrof\'isica de Canarias, E-38205 La Laguna, Tenerife, Spain
  \and
Universidad de La Laguna, Dpto. Astrof\'isica, E-38206 La Laguna, Tenerife, Spain
  \and
Armagh Observatory and Planetarium, College Hill, Armagh, BT61 9DG, UK
  \and
Department of Astronomy, University of California Berkeley, Berkeley, CA 94720, USA}

           \date{Received 27 August 2018 / Accepted 21 December 2018}

           \abstract{We present the first stellar spectroscopy in the
             low-luminosity ($M_V$\,$\sim$\,$-$9.3\,mag), dwarf galaxy
             Leo~P. Its significantly low oxygen abundance (3\% solar)
             and relative proximity ($\sim$1.6\,Mpc) make it a unique
             galaxy in which to investigate the properties of massive
             stars with near-primordial compositions akin to those in
             the early Universe. From our VLT-MUSE spectroscopy we
             find the first direct evidence for an O-type star in the
             prominent H\,{\scriptsize II} region, providing an
             important test case to investigate the potential
             environmental dependence of the upper end of the initial
             mass function in the dwarf galaxy regime. We classify 14
             further sources as massive stars (and 17 more as
             candidate massive stars), most likely B-type objects.
             From comparisons with published evolutionary models we
             argue that the absolute visual magnitudes of massive
             stars in very metal-poor systems such as Leo~P and
             I\,Zw\,18 may be fainter by $\sim$0.5\,mag compared to
             Galactic stars. We also present spectroscopy of two
             carbon stars identified previously as candidate
             asymptotic-giant-branch stars.  Two of three further
             candidate asymptotic-giant-branch stars display
             Ca~{\scriptsize II} absorption, confirming them as cool,
             evolved stars; we also recover Ca~{\scriptsize II}
             absorption in the stacked data of the next brightest 16
             stars in the upper red giant branch.  These discoveries
             will provide targets for future observations to
             investigate the physical properties of these objects and
             to calibrate evolutionary models of luminous stars at
             such low metallicity. The MUSE data also reveal two
             100\,pc-scale ring structures in H$\alpha$ emission, with
             the H~{\scriptsize II} region located on the northern
             edge of the southern ring.  Lastly, we report
             serendipitous observations of 20 galaxies, with redshifts
             ranging from $z$\,$=$\,0.39, to a close pair of
             star-forming galaxies at $z$\,$=$\,2.5.}
\maketitle

\section{Introduction}
\label{intro}

Leo P is a relatively nearby, dwarf irregular galaxy. Its discovery
and first studies of its physical properties were reported in a series
of five papers by \citet{g13}, \citet{r13}, \citet{s13}, \citet{m13},
and \citet {bc14}. Initially discovered from H\,\1 observations from
the Arecibo Legacy Fast ALFA Survey (ALFALFA), further radio
observations revealed its spatial extent and velocity structure
\citep{g13}. Optical imaging from the 3.5\,m WIYN telescope gave a
first view of its stellar population, with evidence of ongoing star
formation in a luminous H\,\2 region \citep{r13}. Spectroscopy of the
H\,\2 region from the 4\,m Mayall Telescope and the Large Binocular
Telescope (LBT) was used to estimate an oxygen abundance, via analysis
of the [O\,\3] \lam4363 line, of [O/H]\,$=$\,7.17\,$\pm$\,0.04
\citep[just 3\% of solar,][]{s13}. We lack an estimate of iron
abundance at present, but even if Leo~P were found to have a sub-solar
$\alpha$/Fe ratio \citep[as found for massive stars in other
metal-poor dwarf irregulars, e.g.][]{v03,tau07,g14,h14}, such a low
oxygen abundance suggests it is one of the most metal-poor
star-forming galaxies known\footnote{Hence the moniker of `Leo P' from
  \citet{g13}, where `P' refers to its pristine nature.}.

Follow-up imaging with the LBT was used by \citet{m13} to estimate a
distance of 1.72\,$^{+0.14}_{-0.40}$\,Mpc, from the location of the
tip of the red giant branch (TRGB) in the colour-magnitude diagram
(CMD), with an improved measurement of 1.62\,$\pm$\,0.15\,Mpc from the
luminosity of the horizontal branch and light curves of candidate
RR\,Lyrae stars \citep{m15}. The system is thought to have a stellar
mass of 5.7\,$\times$\,10$^5$\,M$_\odot$ \citep{m13}, with a
neutral-hydrogen mass of 9.5\,$\times$\,10$^5$\,M$_\odot$ and a
dynamical mass in excess of 2.6\,$\times$\,10$^7$\,M$_\odot$
\citep{bc14}.

Following its discovery, \citet{bofb13} suggested Leo~P is a member of
the NGC\,3109 association and is one of five `dwarfs walking in a row'
(together with NGC\,3109, Sextans~A, Sextans~B, and Antlia).
They speculated this association could have originated from tidal
interaction or, if arriving at the Local Group for the first time,
that the structure could be a remnant of a cosmological filament.

\begin{figure}
\begin{center}
\includegraphics[scale=0.4]{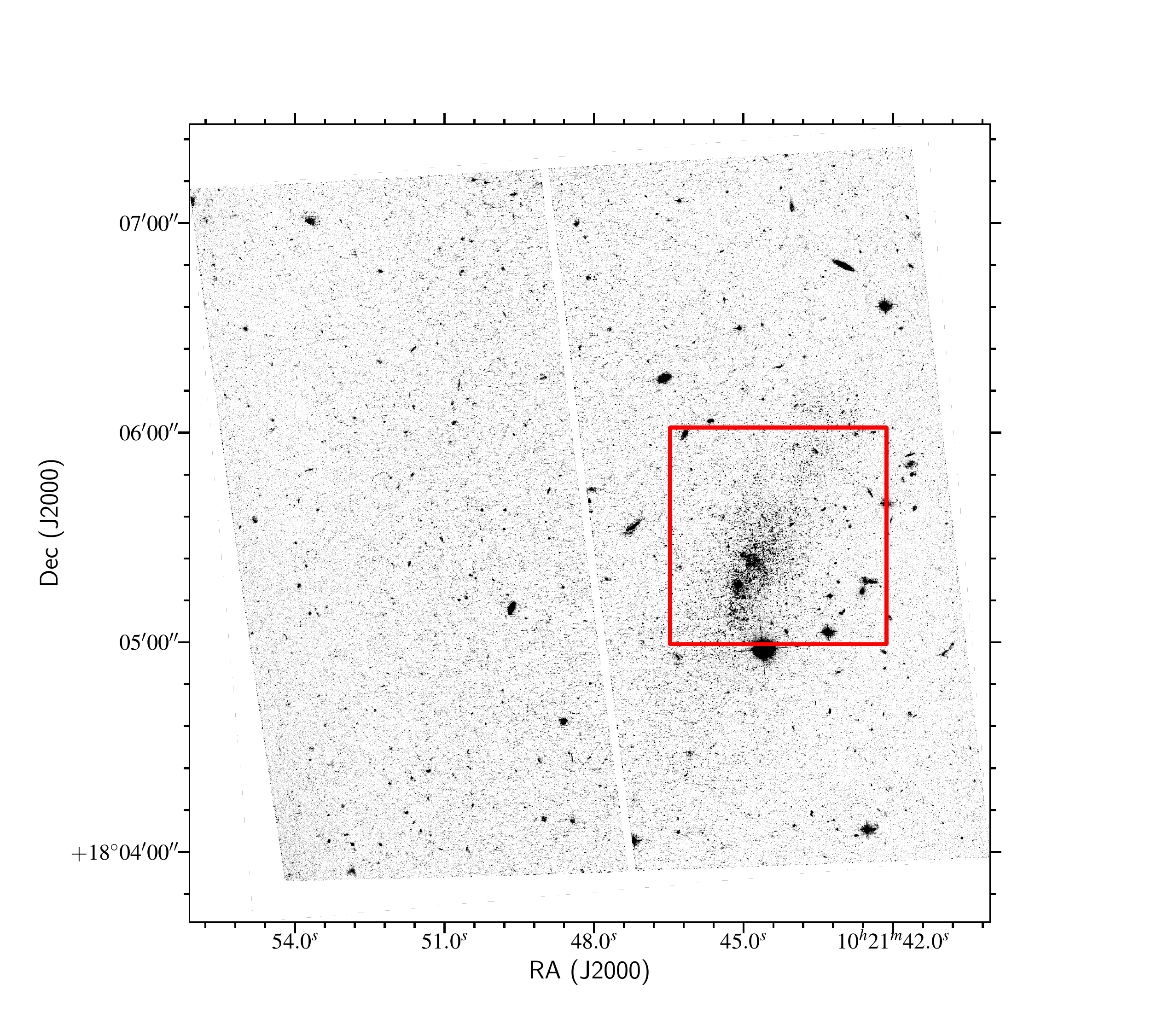}
\caption[]{Observed MUSE field (in red) overlaid on the {\em F475W}
  {\em HST}-ACS image of the Leo~P region \citep{m15}. The MUSE
  data encompass most of the visible extent of this
  low-luminosity dwarf system.}\label{spatial_ACS}
\end{center}
\end{figure}

In this context, Leo P is of interest as the potential site of the
most metal-poor massive stars in the local Universe. 
The populations of high-mass stars in metal-poor galaxies are a prime
contender for the intense ultraviolet radiation which reionised the
intergalactic medium by z\,$\sim$\,6 \citep[e.g.][]{f12,stark16,wbk17}.
However, while there are theoretical predictions for the properties of
high-mass stars at these early epochs (including the putative
`Population III' stars), we lack observational tests of this very
metal-poor regime.

In the Local Group (and nearby), the range of low-metallicity
environments in which to study massive-star evolution is limited to
the Magellanic Clouds \citep[e.g. the VLT-FLAMES
surveys,][]{esl05,vfts} and other metal-poor galaxies such as
IC\,1613, WLM, NGC\,3109 and Sextans~A
\citep[e.g.][]{t11,t14,gh13,c16}. To reach metallicities significantly
below those in the Clouds for substantial stellar populations, we need
to look much further afield to star-forming galaxies such as DDO 68
\citep[$\sim$12.7\,Mpc,][]{c14,s16} or I~Zwicky~18
\citep[18.2\,$\pm$\,1.5\,Mpc,][]{a07}. The presence of young blue
populations in Leo~A and the Sag DIG \citep{weisz14,g18} and the
discovery of Leo~P, gives us a first chance to explore this regime.
Tantalisingly, Leo P has a comparable oxygen abundance to both DDO 68
and I\,Zw\,18 \citep{sk93,p05} but is signficantly closer, potentially
providing a unique opportunity to study high-mass stars in a pristine
environment similar to those in the earliest stages of the Universe.

The cool-star population of Leo~P is of also interest. The main sites
of dust production in the early Universe are thought to be supernovae
\citep[e.g.][]{mmh10} and stars on the asymptotic giant branch
\citep[AGB, e.g.][]{v09,zh13}; luminous blue variables might also play
a role \citep{gn14}. Considerable effort has been undertaken to
identify and characterise the AGB populations of local metal-poor
galaxies to investigate this channel for dust production
\citep[e.g.][]{,boyer09,boyer12,boyer15,boyer17,sloan09,sloan12,j18}. Identification of AGB
stars in Leo~P would be a natural extension to such studies.
\citet{l16} used single-epoch, near-IR imaging of Leo~P to identify 22
candidate AGB stars, but used selection criteria from \citet{s12} from
their observations of NGC\,6822 \citep[where the cool stars are more
metal-rich than Leo~P, e.g.][]{p15}. Spectroscopic confirmation of
their nature is a first step to planning longer-wavelength
observations to investigate dust-production rates.

Here we present the first stellar spectroscopy in Leo~P, obtained with
the Multi-Unit Spectroscopic Explorer \citep[MUSE;][]{b10} on the Very
Large Telescope (VLT) at Paranal. The observations and data reduction
are detailed in Section~2, followed by discussion of the stellar
spectra (and detections of background galaxies) in Section~3. In
Section~4 we discuss the potential impact of the low metallicity of
Leo~P on the magnitudes of massive stars, and in
Section~5 we briefly investigate the morphology and dynamics of the
nebular emission. Concluding remarks are given in Section~6.

\begin{figure*}
\begin{center}
\includegraphics[scale=0.73]{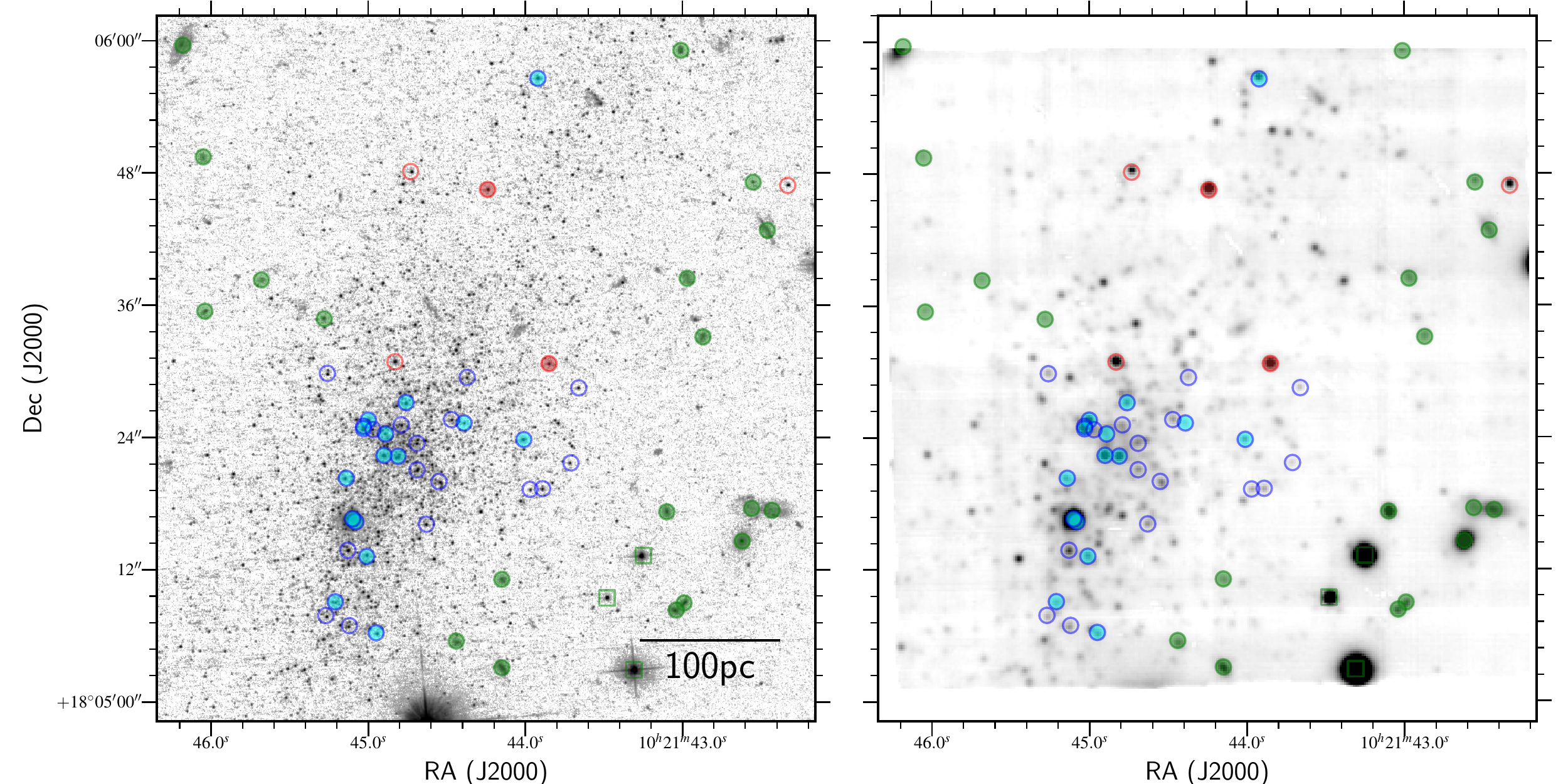}
\caption[]{Spatial distribution of classified spectra in Leo~P. {\it
    Left-hand panel:} {\em HST}-ACS {\em F475W} image; {\it right-hand
    panel:} $V$-band image recovered from the MUSE cube. Objects are
  identified as follows: massive stars (filled cyan circles),
  candidate massive stars (open blue circles), carbon stars (filled
  red circles), additional candidate AGB stars from \citet[][open red
  circles]{l16}, background galaxies (green circles), foreground stars
  (open green squares). The 100\,pc scalebar assumes the distance of
1.62\,Mpc from \citet{m15}.}\label{spatial}
\end{center}
\end{figure*}

\section{Observations \& data reduction}
We observed Leo~P using the extended wide-field mode of MUSE, without
the blue filter to extend the wavelength coverage down to 4650\,\AA\/
(to include He~\2 \lam4686 if present in our targets). Eight
observations, spanning 13 December 2015 to 8 March 2016, were executed
from the service queue on UT4 of the VLT. Each comprised
2\,$\times$\,1505\,s exposures (so a total integration of 6.7\,hrs),
centred on a position of $\alpha$\,$=$\,10$^{\rm h}$\,21$^{\rm
  m}$\,44{\mbox{\ensuremath{.\!\!^{\,\rm s}}}}3,
$\delta$\,$=$\,18$^\circ$\,05$'$\,30\farcs5 (J2000). As shown in
Fig.~\ref{spatial_ACS}, the observed MUSE field spans most of the
visible extent of the galaxy. All of the observations were obtained in
dark time (in terms of lunar illumination and angular distance),
except for one that was on the boundary of grey and bright time.

The data were reduced using the MUSE pipeline (v1.6.1), with steps
including: bias correction, wavelength calibration, reconstruction of
datacubes from the individual spectra on the detectors, correction of
the spectra to the heliocentric frame, sky subtraction, and merging of
the individual exposures to form a combined datacube. Our reductions
were largely indistinguishable to the `MUSE-DEEP' processing and
stacking of the data undertaken as part of ESO's archiving
activities, and the combined cube is available from the ESO Phase~3
archive\footnote{http://archive.eso.org/wdb/wdb/adp/phase3\_main/form}.
The central parts of the cube are relatively crowded (cf. the typical
seeing of the observations of $\sim$0\farcs6). To extract our sources
we used the {\sc pampelmuse} software \citep{k13} that was developed
to recover spectra from blended sources in MUSE data \citep[see,
e.g.][]{k16}.

For the input catalogue for the {\sc pampelmuse} extractions we used
the {\em Hubble Space Telscope (HST)} imaging from \citet{m15}, who
used the Advanced Camera for Surveys (ACS) to observe Leo~P with the
{\em F475W} and {\em F814W} filters (see Fig.~\ref{spatial_ACS}). We
created a source catalogue from the {\em HST} data with {\sc dolphot},
a version of {\sc hstphot} \citep[][that has been updated with a
specific module for ACS]{d00}, and using the parameter file and
photometric criteria from \citet{w14}. The angular resolution of these
data is a factor of ten better than the MUSE observations, and many of
the MUSE sources are resolved into multiple components.

{\sc pampelmuse} is designed to fit the point spread function (PSF) of
each source throughout each wavelength slice of the combined MUSE
datacube. In general, this requires a subset of bright, high
signal-to-noise stars to define the PSF and coordinate transformations
between the input catalogue and the MUSE observations prior to the
extractions. Unfortunately, our Leo~P field lacked sufficiently bright
stars for adequate PSF modelling. This is in contrast to the recent
analysis of the luminous stellar population of NGC\,300 which,
although more distant at $\sim$1.9\,Mpc, has a much greater stellar
density and a significant population of luminous stars \citep{r18}. 

We therefore used {\sc pampelmuse} to fit all resolved sources
simultaneously, which entailed computing both the Gaussian PSF profile
and the coordinate transformations at each layer. After {\sc
  pampelmuse} had selected all the stars that were suitable for
fitting and extraction, we performed a first run using a binning of 50
pixels in the spectral direction. This allowed us to smooth the
wavelength dependencies to the fitted parameters by using a simple
polynomial fit before performing the final extraction at each
wavelength for 341 sources from the {\em HST} catalogue. In the cases
of well isolated stars the resulting spectra were comparable to simple
aperture extractions from the cube. However, the {\sc pampelmuse}
approach has the advantage over aperture extractions of not simply
co-adding multiple components together in one spectrum (where objects
are either blended or simply within the defined aperture).

The positions and photometry of the MUSE spectra that we were able to
classify are given in Tables~\ref{targets} and \ref{cool_targets}. We
also give the identifiers and photometry from the ground-based imaging
from \citet[MSB,][]{m13}. Those data were obtained with a seeing of
0\farcs7, so are a good match to the angular resolution of our MUSE
cube for relatively isolated stars (or where additional components
detected in the {\em HST} images are sufficiently faint not to
contribute more than a few percent of the flux). Given the small field
of the ACS images, the right ascension and declination of our sources
are given in the astrometric frame of \citet{m13}, where the {\em HST}
and MUSE data were transformed using matched stars from their
catalogue.

\section{Spectral content}
\begin{figure}
\begin{center}
\includegraphics[scale=0.75]{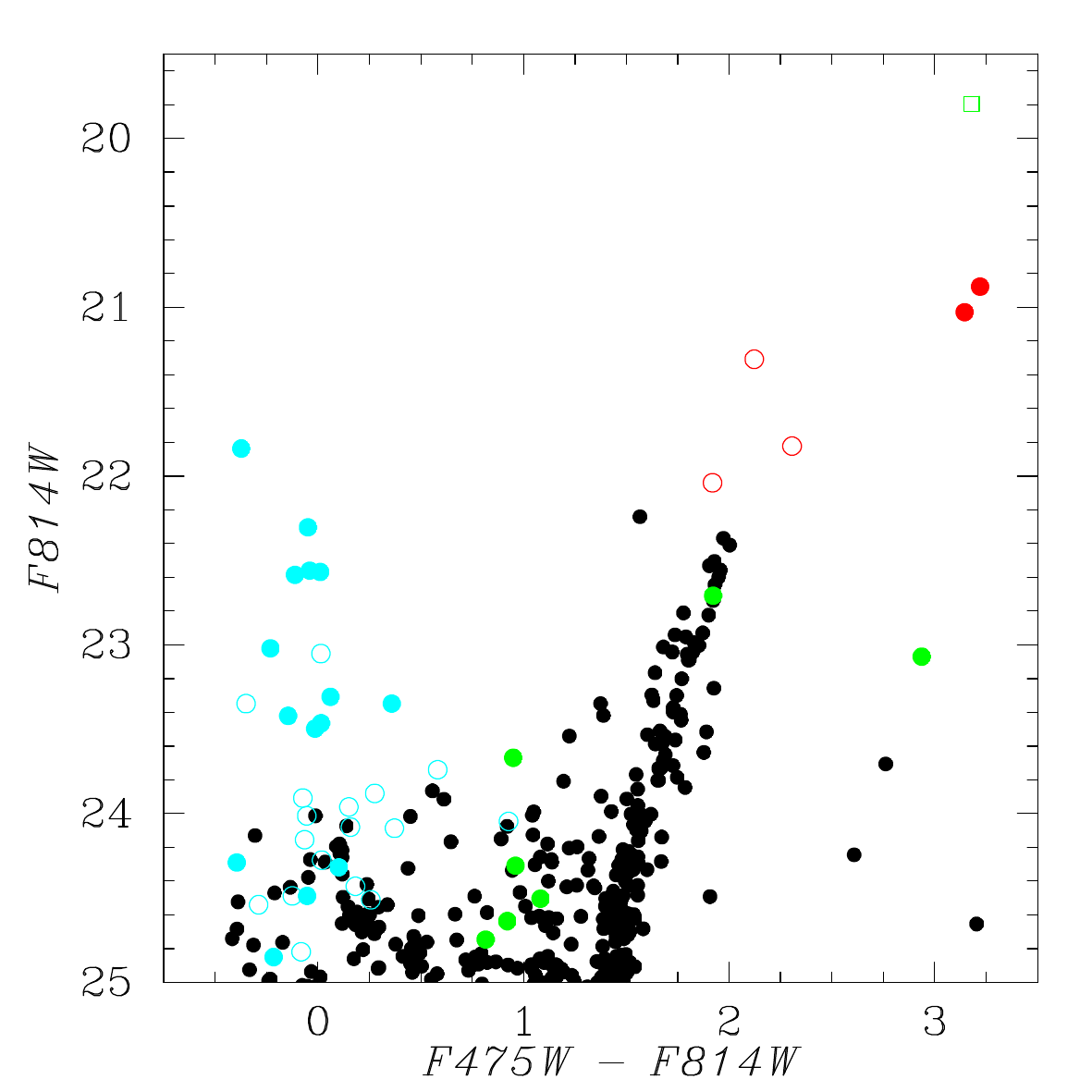}
\caption[]{Colour-magnitude diagram for the MUSE field using our 
  photometry of the {\em HST} imaging from \citet{m15}. Symbols as
  follows: massive stars (closed cyan circles), candidate massive stars (open
  cyan circles), carbon stars (closed red circles), other sources
  previously classified as potential AGB stars (open red circles),
  background galaxies (green circles), foreground star (open green
  square).}\label{cmd}
\end{center}
\end{figure}

Our motivation for the MUSE observations was a first spectroscopic
census of Leo~P, primarily to identify candidate massive stars for
quantitative analysis with higher-quality, follow-up spectroscopy.
The spatial locations of our classified spectra are shown in
Fig.~\ref{spatial}, with their positions in the {\em HST} CMD shown in
Fig.~\ref{cmd}. For reference, \citet{m13} estimated the TRGB as
$I$\,$=$\,22.11\,mag. As such, we expected to find a number of
(modestly) massive stars in the MUSE spectroscopy in the `blue' plume
at ($F475W$\,$-$\,$F814W$)\,$\sim$\,0.0\,mag, and luminous evolved red
stars above the TRGB. We have found examples of both hot and cool
objects, as well as background galaxies in the MUSE field, as outlined
in the following sections.

\begin{figure}
\begin{center}
\includegraphics[scale=0.75]{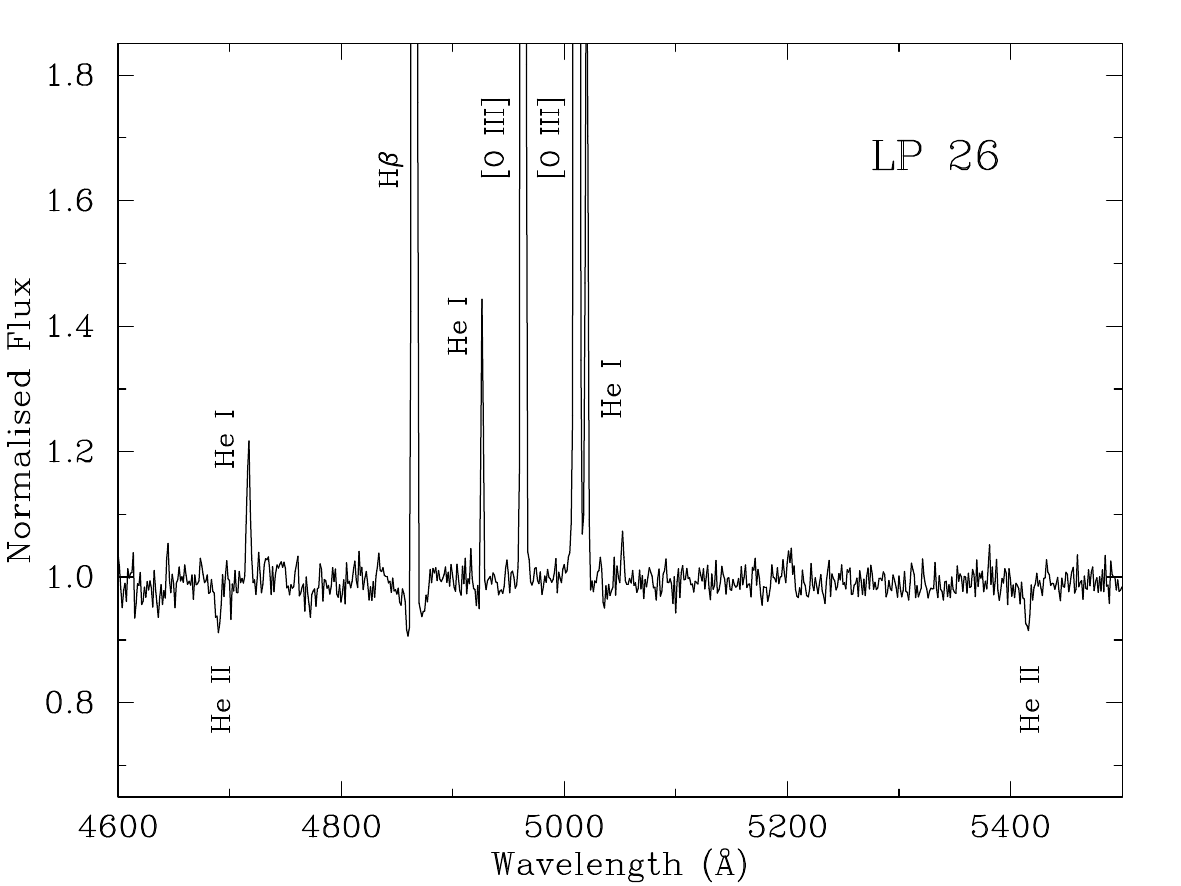}
\caption[]{Blueward part of the MUSE spectrum of the extracted central
  source of the H\,{\scriptsize II} region in Leo~P (LP\,26). The identified
  emission lines, from left-to-right by species, are: H$\beta$;
  He~{\scriptsize I} \lam\lam4713,4922,5016; [O~{\scriptsize III}]
  \lam\lam4959,5007. The He~{\scriptsize II} \lam\lam4686, 5411
  absorption lines provide the first direct evidence for an O-type
  star in Leo~P.}\label{HIIregion}
\end{center}
\end{figure}

\subsection{HII region}
There is a prominent H~\2 region in Leo~P, discussed by \citet{s13},
and resolved into multiple sources and nebulosity by the {\em HST}
images \citep{m15}. Skillman et al. mentioned an absence of absorption
features, but the MUSE spectrum of the bright, central star (LP\,26)
displays He~\2 \lam\lam4686, 5411 absorption (see
Fig.~\ref{HIIregion}), arguing for the presence of an O-type star.
\citet{m15} estimated the absolute magnitude of this source as
$M_V$\,$=$\,$-$4.43, equivalent to what we would expect for a mid to
late O-type dwarf at Galactic metallicity \citep[e.g.][]{w72,msh05}.
In the absence of published calibrations for the absolute magnitudes
of O-type stars at low metallicities, we discuss this aspect further
in Section~\ref{mags}.

\begin{table*}
\begin{center}
\caption{Confirmed and candidate blue stars in the Leo~P MUSE field.}\label{targets}
{\tiny
\begin{tabular}{llcccccccl}
  \hline\hline
LP\# & MSB   & $\alpha$\,(J2000) & $\delta$\,(J2000) & $V$ & $I$ & {\em F475W}&{\em F814W} & S/N & Comments\\
  \hline
1  &  138  & 10 21 43.66 & $+$18 05 28.50 & 24.40 & 23.90 & 24.459 & 24.087 & \o4 & candidate \\
2  &  207  & 10 21 43.71 & $+$18 05 21.69 & 24.79 & 24.64 & 24.737 & 24.819 & \o5 & candidate \\
3  &  94   & 10 21 43.89 & $+$18 05 19.35 & 23.98 & 23.68 & 24.156 & 23.880 & \o7 & candidate \\
4  &  16   & 10 21 43.92 & $+$18 05 56.59 & 22.50 & 22.29 & 22.521 & 22.561 &  19 & RV(H$\beta$)\,$=$\,240\,$\pm$\,19\,\kms \\
5  &  202  & 10 21 43.97 & $+$18 05 19.29 & 24.83 & 24.09 & 24.973 & 24.046 & \o4 & candidate \\
6  &  40   & 10 21 44.01 & $+$18 05 23.81 & 23.45 & 23.08 & 23.707 & 23.348 & \o8 & \\
7  &  98   & 10 21 44.37 & $+$18 05 29.45 & 24.09 & 23.52 & 24.613 & 24.431 & \o5 & candidate \\
8  &\ldots & 10 21 44.39 & $+$18 05 25.31 &\ldots &\ldots & 24.419 & 24.318 & \o9 & \\
9  &  92   & 10 21 44.47 & $+$18 05 25.62 & 23.95 & 23.66 & 24.111 & 23.961 & \o6 & candidate \\
10 &\ldots & 10 21 44.55 & $+$18 05 20.00 &\ldots &\ldots & 23.958 & 24.013 & \o9 & candidate \\
11 &\ldots & 10 21 44.63 & $+$18 05 16.14 &\ldots &\ldots & 23.834 & 23.908 & \o4 & candidate \\
12 &\ldots & 10 21 44.69 & $+$18 05 21.07 &\ldots &\ldots & 24.363 & 24.488 & \o7 & candidate \\
13 &\ldots & 10 21 44.69 & $+$18 05 23.47 &\ldots &\ldots & 24.090 & 24.155 & \o9 & candidate \\ 
14 &  25   & 10 21 44.76 & $+$18 05 27.17 & 23.14 & 23.00 & 23.276 & 23.421 &  10 & Be/Ae? (H$\alpha$ em.?)\\
15 &\ldots & 10 21 44.79 & $+$18 05 25.15 &\ldots &\ldots & 24.251 & 24.540 & \o7 & candidate \\
16 &  15   & 10 21 44.81 & $+$18 05 22.31 &\ldots &\ldots & 22.790 & 23.021 &  21 & \\
17 &\ldots & 10 21 44.89 & $+$18 05 24.33 &\ldots &\ldots & 24.634 & 24.850 & \o8 & \\
18 &  12   & 10 21 44.90 & $+$18 05 22.38 & 22.16 & 22.02 & 22.255 & 22.304 &  28 & RV(H$\beta$)\,$=$\,274\,$\pm$\,19\,\kms \\
19 &  41   & 10 21 44.95 & $+$18 05 06.28 & 23.46 & 23.37 & 23.483 & 23.497 & \o8 & \\
20 &\ldots & 10 21 44.97 & $+$18 05 24.72 &\ldots &\ldots & 24.768 & 24.511 & \o6 & candidate \\
21 &\ldots & 10 21 45.00 & $+$18 05 25.59 & \ldots& \ldots& 22.579 & 22.568 &  15 & RV(H$\beta$)\,$=$\,260\,$\pm$\,22\,\kms \\
22 &  24   & 10 21 45.01 & $+$18 05 13.21 & 23.15 & 22.87 & 23.369 & 23.308 &  11 & Be/Ae? (H$\alpha$ \& H$\beta$ em.?)\\
23 &\ldots & 10 21 45.03 & $+$18 05 24.80 & \ldots& \ldots& 22.473 & 22.585 &  12 & RV(H$\beta$)\,$=$\,247\,$\pm$\,20\,\kms \\
24 &\ldots & 10 21 45.03 & $+$18 05 25.02 & \ldots& \ldots& 23.066 & 23.052 & \o3 & candidate \\
25 &\ldots & 10 21 45.08 & $+$18 05 16.35 & \ldots& \ldots& 24.290 & 24.684 &  15 & On SE edge of H~{\scriptsize II} region\\
26 &  2    & 10 21 45.10 & $+$18 05 16.62 & \ldots& \ldots& 21.464 & 21.837 &  52 & O-type star in H~{\scriptsize II} region\\
27 &  82   & 10 21 45.12 & $+$18 05 06.92 & 23.92 & 23.36 & 24.322 & 23.740 & \o5 & candidate \\
28 &  22   & 10 21 45.13 & $+$18 05 13.76 & 23.01 & 23.16 & 22.999 & 23.349 &  13 & candidate \\
29 &  33   & 10 21 45.14 & $+$18 05 20.31 & 23.27 & 23.09 & 23.480 & 23.465 & \o7 & \\ 
30 &  63   & 10 21 45.21 & $+$18 05 09.09 & 23.73 & 23.41 & 24.435 & 24.488 & \o8 & \\
31 &  114  & 10 21 45.26 & $+$18 05 29.82 & 24.17 & 23.89 & 24.294 & 24.276 & \o4 & candidate \\
32 &  106  & 10 21 45.27 & $+$18 05 07.84 & 24.11 & 23.90 & 24.238 & 24.080 & \o5 & candidate \\ 
\hline
\end{tabular}
\tablefoot{Identifiers in the second col. and $V$- and
  $I$-band photometry are from \citet[MSB,][]{m13}; {\em F475W} and
  {\em F814W} photometry is from the {\em HST} imaging from
  \citet{m15}. S/N estimates (per pixel) were obtained over the
  5050-5400\,\AA\ part of the normalised spectra.}}
\medskip
{\tiny
\caption[]{Confirmed and candidate cool stars in the MUSE field.\label{cool_targets}}
\begin{tabular}{lccccccccll}
\hline\hline
MSB  & $\alpha$\,(J2000) & $\delta$\,(J2000) & $V$ & $I$ &{\em F475W}&{\em F814W}& $J$ & $K$ & Classification & Comments \\
\hline
18  & 10 21 42.33 & $+$18 05 46.87 & 23.27 & 21.70 & 24.129 & 21.823 & 20.972 & 20.254 & AGB?          & Cand. M-AGB (L16) \\
10  & 10 21 43.85 & $+$18 05 30.70 & 23.00 & 20.97 & 24.175 & 21.029 & 20.073 & 18.773 & Carbon star   & Cand. C-AGB (L16) \\
8   & 10 21 44.24 & $+$18 05 46.50 & 22.97 & 20.90 & 24.099 & 20.878 & 19.464 & 18.451 & Carbon star   & Cand. M-AGB (L16) \\
13  & 10 21 44.83 & $+$18 05 30.87 & 22.63 & 21.20 & 23.430 & 21.308 & 20.495 & 19.691 & AGB?          & Cand. M-AGB (L16) \\ 
21  & 10 21 44.73 & $+$18 05 48.13 & 23.30 & 21.97 & 23.959 & 22.040 & 21.198 & 20.311 & ?             & Cand. M-AGB (L16) \\
\hline 
1 & 10 21 43.27 & $+$18 05 13.36 & 20.02 & 20.03 & \ldots & \ldots & 15.692 & 14.979 & M (f/g?) & 2MASS:\,J10214334$+$1805133\\ 
\ldots& 10 21 43.32 & $+$18 05 03.22 &\ldots &\ldots & \ldots & \ldots & 16.156 & 15.362 & K (f/g)  & 2MASS:\,J10214332$+$1805032 \\ 
5   & 10 21 43.48 & $+$18 05 09.43 & 22.05 & 20.27 & 22.975 & 19.796 & \ldots & \ldots & M (f/g)  & \\ 
\hline
\end{tabular}
\tablefoot{Identifiers and $V$- and $I-$band photometry are from
  \citet{m13}; {\em F475W} and {\em F814W} photometry is from {\em
    HST} imaging by \citet{m15}. IR photometry is from \citet[][with
  his classifications in the final col.]{l16}, and 2MASS \citep{2mass}
  for two of the foreground objects.}}
\end{center}
\end{table*}

\subsection{Early-type stars}

At the spectral resolution ($R$\,$\sim$\,2000 at 5500\,\AA) and
signal-to-noise (S/N) of the MUSE data, we would not expect to be able
to identify many prominent spectral lines in early-type stars (which
is further exacerabated by the lack of spectral coverage shortwards of
$\sim$4650\,\AA). Nonetheless, on the basis of H$\beta$ absorption,
combined with indications of stellar features at H$\alpha$, we
classify 14 sources as massive stars, with spectra of some of the
brighter sources shown in Fig.~\ref{blue} (with a S/N ranging from 12
to 28 per pixel). We classified a further 17 spectra as possible
massive stars, typically where the H$\beta$ absorption is less secure,
but still supported by an increasingly blue flux distribution.

Given this data quality, combined with the spectral range, sky
residuals, and resolution of MUSE, it is not surprising that we do not
see other strong features. That said, weak absorption consistent with
arising from He~\1 \lam\lam5876, 6678 is seen in both LP\,16 and LP\,18 (see
Fig.~\ref{blue}), but we also see evidence for oversubtraction of the
[O~\3] nebular lines (and similarly in the core of H$\alpha$)
suggesting these could be related to limitations of the background
subtraction. The only other feature of note is Paschen\,9 \lam9229
absorption in the spectra of LP\,21 and LP\,23.
From the steepening blue flux distributions of these spectra, combined
with an absence of features that might be expected at cooler types, we
expect that most of these are B-type objects (or potentially late
O-type for the brightest few objects). The location of these sources
in the CMDs (Fig.~\ref{cmd}) provides support for our classifications
in most cases.

For four spectra with S/N\,$>$\,10 and well-defined H$\beta$
absorption (and without obvious evidence for nebular over-subtraction)
we estimated radial velocities (RV) from Gaussian fits to the observed
H$\beta$ absorption (included in last column of Table~\ref{targets}).
The weighted-mean velocity ($\overline{\rm RV}$) from the four spectra
is $\overline{\rm RV}$\,$=$\,255\,$\pm$\,13\,\kms. The velocity of
the He~\2 \lam4686 absorption in the spectrum of LP\,26 in the H~\2
region is consistent with these values (252\,$\pm$\,35\,\kms). Given the
relatively large uncertainties on the MUSE values, they
are in good agreement with the systemic velocity of
neutral hydrogen of $v_{\rm LSRK}$\,$=$\,260.8\,$\pm$\,2.5\,\kms\/
\citep{bc14}, which equates to a heliocentric velocity of approx.
265\,\kms\/ in the direction of Leo~P.

\subsection{Cool stars}
\begin{figure*}
\begin{center}
\includegraphics[scale=0.9]{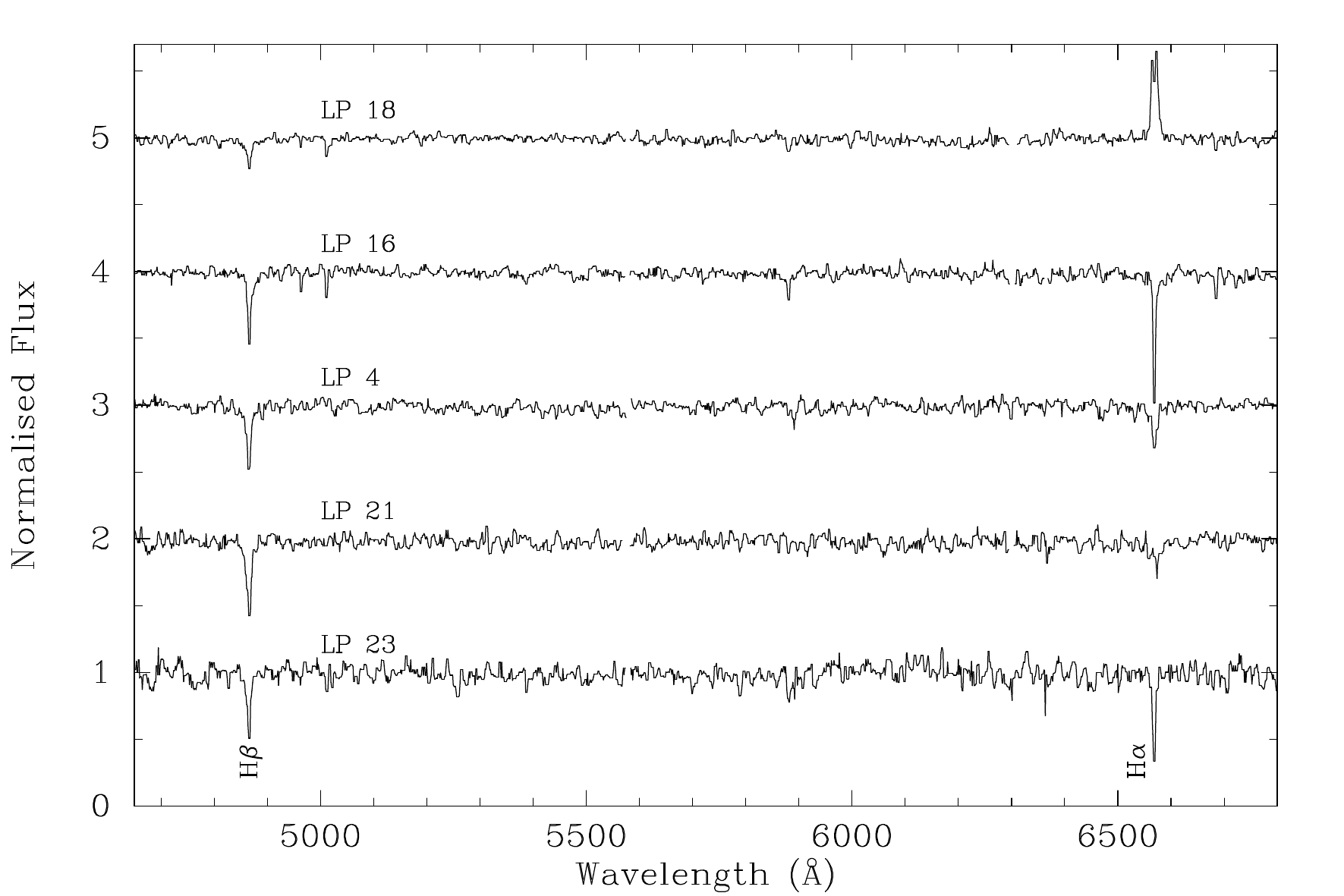}
\caption[]{Best five examples of spectra of presumed BA-type stars
  in Leo~P (each smoothed using a five-pixel boxcar filter). In addition
  to the Balmer line absorption, there is tentative detection of
  He~{\scriptsize I} $\lambda\lambda$5876, 6678 absorption in LP\,16 and
  LP\,18, although potentially related to the apparent over-subtraction
  of nebular features (as demonstrated by the [O~{\scriptsize III}]
  $\lambda\lambda$4959, 5007 `absorption' and similar evidence in the
  cores of the H$\alpha$ lines).}\label{blue}
\medskip
\includegraphics[scale=0.9]{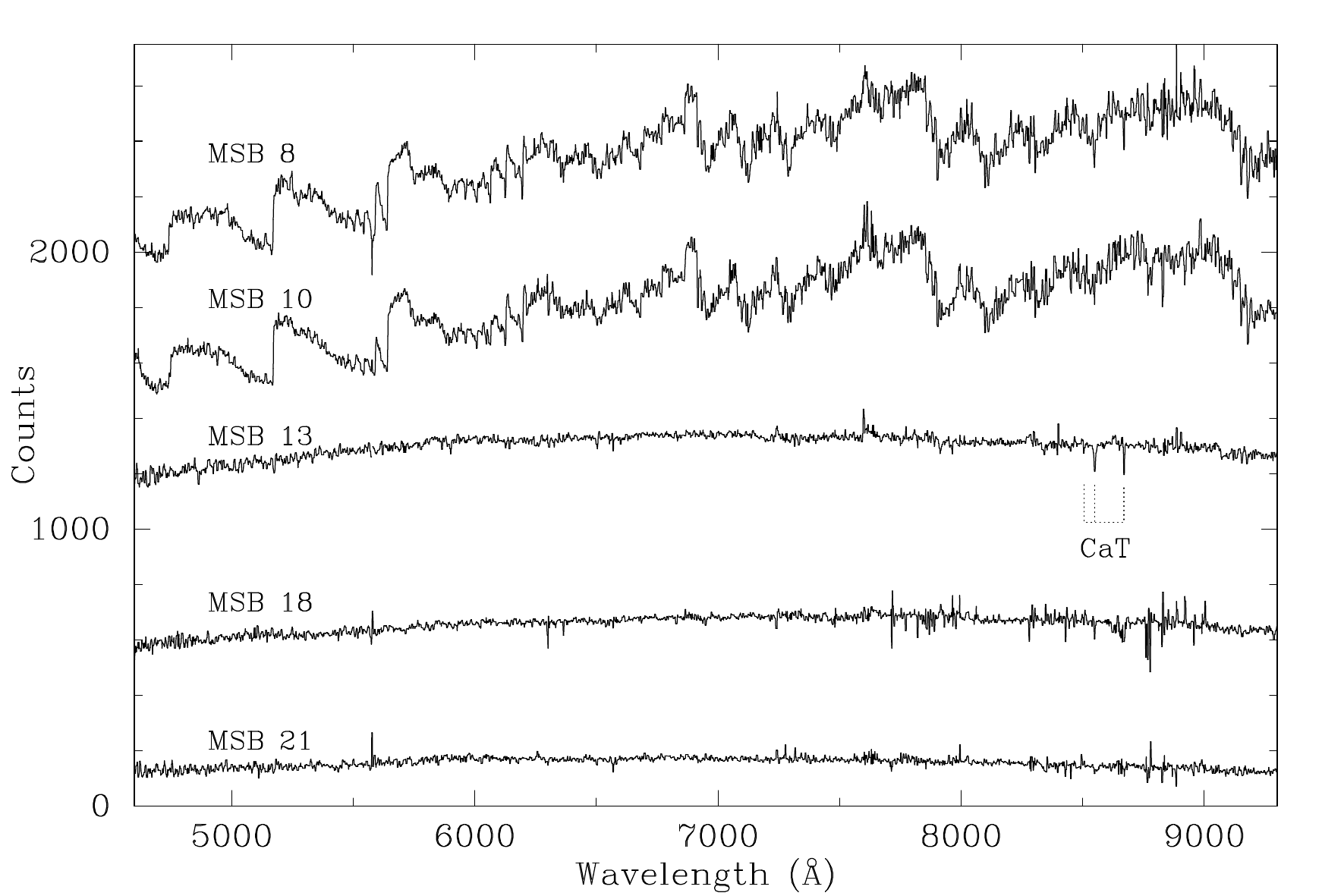}
\caption[]{MUSE spectra of candidate AGB stars from \citet{l16}, in
  which each spectrum was smoothed using a five-pixel boxcar filter and
  displaced vertically by increasing multiples of 500 counts. The
  expected location of absorption by the Ca~{\scriptsize II} triplet
  is indicated for MSB\,13, with an expanded view of this region shown
  in Fig.~\ref{CaT}. Mean counts over the 8110-8270\,\AA\/ range
  (which is relatively free of sky residuals) for the five spectra
  are, from top to bottom, 396, 343, 314, 171, 125, in keeping with
  the $I$-band and {\em F814W} magnitudes in
  Table~\ref{cool_targets}.}\label{agb_lee}
\end{center}
\end{figure*}

Near-infrared imaging of Leo~P was used by \citet{l16} to identify 22
candidate AGB stars, six of which lie within the MUSE field. The
spectra of five of these are shown in Fig.~\ref{agb_lee}, with their
observational properties summarised in Table~\ref{cool_targets}; the
sixth (MSB\,\#56) is a background galaxy (see
Table~\ref{highz_targets}). The two brightest candidates are carbon
stars, as revealed by the strong C$_2$ Swan bandheads at \lam\lam5165,
5636. We compared the wavelengths of the bandheads in the MUSE spectra
to those in an archival UVES spectrum from \citet{uvespop} of the
Galactic carbon star, W~Ori (HD\,32736), finding differential
velocities that are consistent with their membership of Leo~P (i.e.
$\delta$RV\,$\sim$\,250\,\kms).

\begin{figure}
\begin{center}
\includegraphics[scale=1.0]{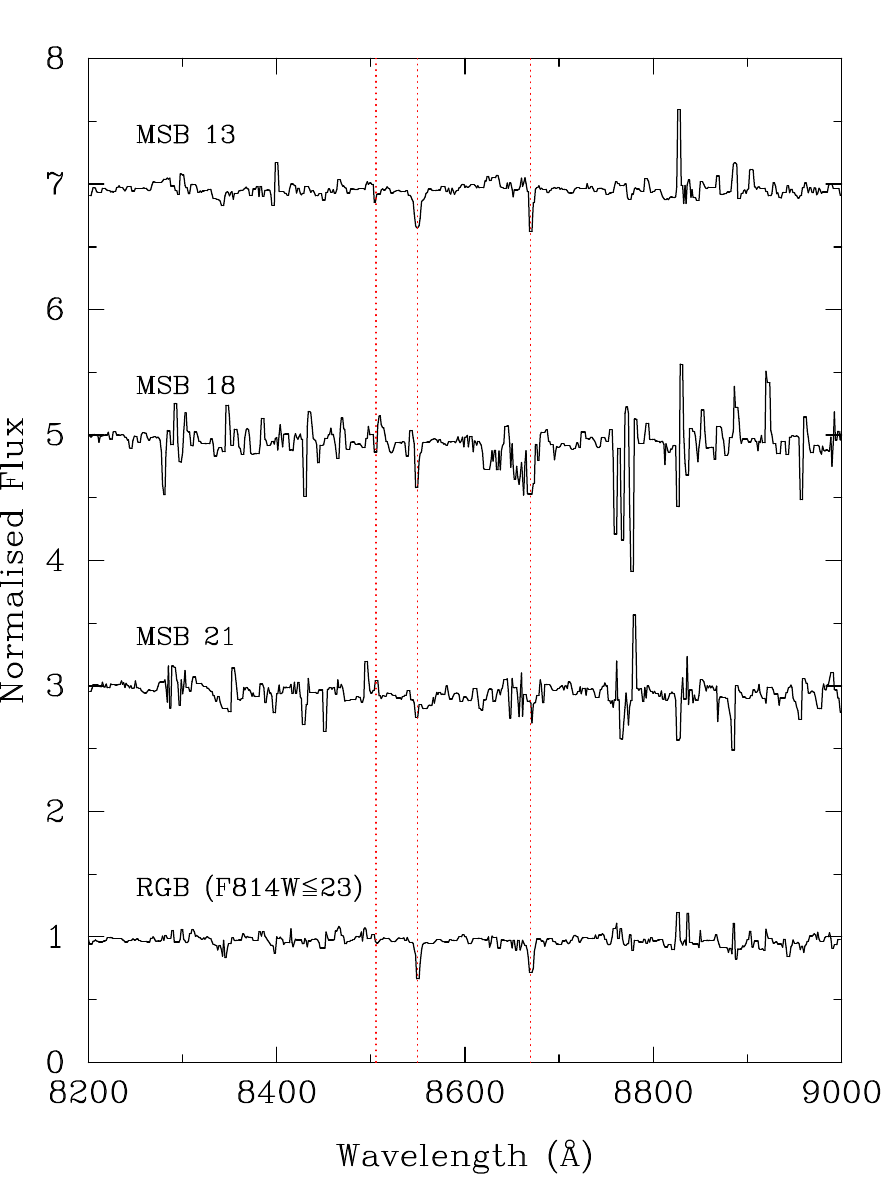}
\caption{Region around the Ca~{\scriptsize II} triplet (CaT) for the
  three stars previously identified as candidate AGB stars, together
  with a stacked spectrum of the 16 next brightest objects at the top of
  the red giant branch (i.e. excluding the three candidate AGB stars). CaT
  absorption is seen in the brightest two (MSB\,13 and 18) and in the 
  stacked data.}\label{CaT}
\end{center}
\end{figure}

The remaining three stars, suggested as oxygen-rich M-type AGB stars
by \citeauthor{l16}, are relatively featureless. However, there
appears to be absorption in the near-IR Ca~\2 Triplet (CaT, with rest
wavelengths of \lam\lam8498, 8542, 8662) for MSB\,13. An expanded view
of the CaT region for the candidate AGB sources is shown in
Fig.~\ref{CaT}, with absorption also present for \lam8542 in MSB\,18.
Indeed, of the three CaT lines in the MUSE data, the central \lam8542
line is the most robust in terms of being less influenced by sky
residuals. From a Gaussian fit to the \lam8542 line in MSB\,13 we
estimated RV\,$=$\,262\,$\pm$\,8\,\kms, in good agreement with the
values for the hotter stars. To investigate the CaT region for the
fainter stars in the RGB, the lower spectrum in Fig.~\ref{CaT} shows
the co-added data of the 16 sources with {\em F814W}\,$\le$\,23.0.
The stronger CaT components can again be seen, confirming the presence
of cool, evolved stars as expected from the CMD (Fig.~\ref{cmd}).

Thus, we conclude that at least two of these candidate AGB stars
appear to be cool, evolved stars with CaT absorption (and with $F814W$
magnitudes that are brighter than the TRGB). The corresponding lack of
strong molecular bands (e.g. from TiO) in their spectra in
Fig.~\ref{agb_lee} is notable compared with what is usually seen in
(M-type) AGB specta, but not unexpected given the low metallicity of
Leo~P (as traced by the published oxygen abundance). Indeed, the
ratio of C-rich to M-type (O-rich) AGB stars is known to be a function
of metallicity, with an increasing C/M ratio towards lower
metallicities \citep[e.g.][]{ir83,bd05}. The relative dearth of M-type
objects at low metallicities is attributed to the quicker dredge-up
timescales to become a carbon star, combined with higher temperatures
from the evolutionary models, which would act to reduce the TiO
absorption, hence fewer M-type spectra \citep[see, e.g., discussion
by][]{k12}. Equally, it is plausible that simple metallicity effects
give the impression of earlier spectral types, even if the
temperatures were not that different. For example, a luminous, cool
star with an effective temperature of 4000\,K would be classified as
M-type in the LMC, but at the metallicity of Leo~P, the TiO bands
would weaken sufficiently that it would be classified as a K-type
spectrum.

All five of the objects from \citet{l16} are relatively isolated,
point-like sources in the {\em HST} imaging, and we highlight that
four are somewhat removed from the central part of the system. This
suggests the intermediate-age population is either more extended or
well mixed, with the latter similar to the findings for six of the
nine dwarf galaxies studied by \citet{m17}. Following the arguments
of \citet{m17}, we note that MSB\,18 is at a distance of 50$''$
(equivalent to 400\,pc) from the H~\2 region, potentially providing a
source of chemical enrichment in the outer regions of Leo~P.

\subsection{Foreground stars}
There are three seemingly foreground stars included in our extractions
from the MUSE cube, as listed at the end of Table~\ref{cool_targets}.
One star (MSB\,1) has a somewhat larger RV than might be expected for
a foreground Galactic object. Cross-correlation of the two M-type
spectra (over the \lam\lam4600-5565 range) yielded a differential RV
of 168\,$\pm$\,20\,\kms, in good agreement with absolute estimates
from the CaT of 172$\,\pm$\,6\,\kms.

\subsection{Background galaxies}\label{highz_sources}
\begin{figure}
\begin{center}
\includegraphics[scale=0.35]{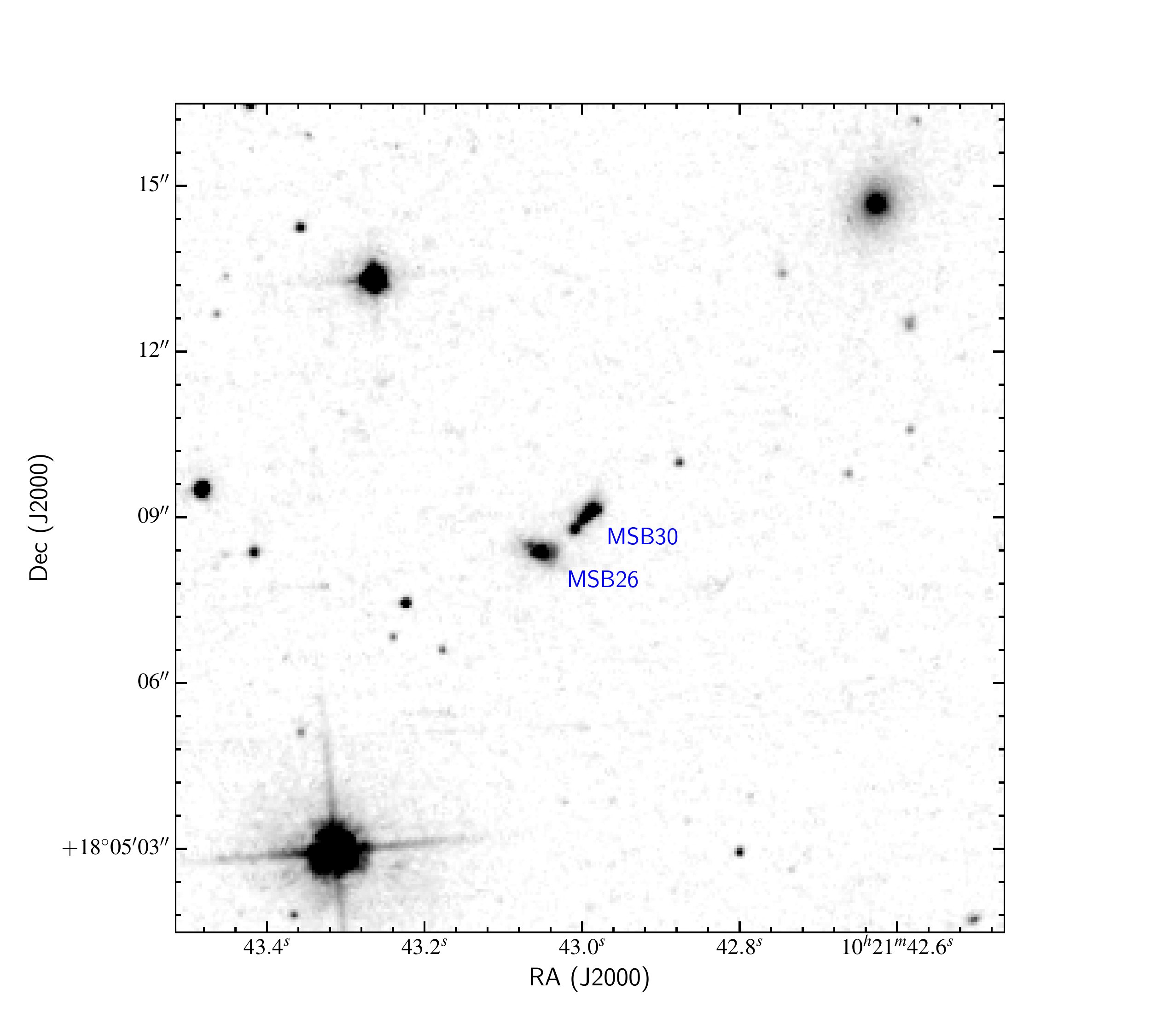}
\caption[]{Two $z$\,$\sim$\,2.5 sources in the {\em HST}-ACS {\em F475W}
  image from \citet{m15}. At the resolution of {\em HST}, the
  extracted MUSE sources are both resolved into two
  components.}\label{z25_hst}
\end{center}
\end{figure}

From initial extractions of the MUSE sources using source detections
and simple aperture extractions, we found 20 sources which are
background galaxies. Indeed, a number of these are clearly visible as
such in the {\em HST} imaging \citep[see discussion of two examples
by][]{l16}, meaning that they are not included in the CMD in
Fig.~\ref{cmd} as they were rejected as non-stellar sources in the
photometric analyis. Observational details for the
spectroscopically-confirmed galaxies are given in
Table~\ref{highz_targets}, including estimates of their redshifts
($z$) using the diagnostic emission or absorption features as indicated.
Of note are the two systems at $z$\,$=$\,2.5, located adjacent to each
other on the sky. The {\em HST} imaging reveals that both MUSE sources
are comprised of two components, as
shown in Fig.~\ref{z25_hst}.

\begin{table*}
\begin{center}
\caption[]{Background galaxies in the MUSE field.}\label{highz_targets}
{\tiny
\begin{tabular}{lcccccl}
\hline\hline
MSB   & $\alpha$\,(J2000) & $\delta$\,(J2000) & $V_o$ & $I_o$ & $z$ & Diagnostic \\
\hline
19    & 10 21 42.43 & $+$18 05 17.38 & 23.41 & 21.70 & 0.5436 & [O~{\scriptsize II}] \lam3727 \\
39    & 10 21 42.46 & $+$18 05 42.80 & 23.88 & 22.33 & 0.7543 & [O~{\scriptsize II}] \lam3727 \\
56    & 10 21 42.55 & $+$18 05 47.16 & 24.62 & 22.53 & 0.5169 & [O~{\scriptsize II}] \lam3727 \\
37    & 10 21 42.56 & $+$18 05 17.56 & 24.19 & 22.24 & 0.7554 & [O~{\scriptsize II}] \lam3727 \\
4     & 10 21 42.62 & $+$18 05 14.59 & 21.95 & 20.26 & 0.3925 & Ca~$H+K$ \\
62    & 10 21 42.87 & $+$18 05 33.13 & 24.15 & 22.64 & 0.7539 & [O~{\scriptsize II}] \lam3727 \\
48    & 10 21 42.97 & $+$18 05 38.43 & 24.22 & 22.40 & 0.7549 & [O~{\scriptsize II}] \lam3727 \\
30    & 10 21 42.99 & $+$18 05 08.98 & 23.30 & 22.62 & 2.5007 & Near-UV Fe~{\scriptsize II}, Mg~{\scriptsize II} \\
49    & 10 21 43.01 & $+$18 05 59.13 & 23.98 & 22.50 & 0.8883 & [O~{\scriptsize II}] \lam3727 \\
26    & 10 21 43.04 & $+$18 05 08.34 & 23.29 & 22.36 & 2.5005 & Near-UV Fe~{\scriptsize II}, Mg~{\scriptsize II} \\
7     & 10 21 43.10 & $+$18 05 17.26 & 23.07 & 20.88 & 0.5440 & [O~{\scriptsize II}] \lam3727 \\
31    & 10 21 43.39 & $+$18 05 38.07 & 23.60 & 22.25 & 0.7462 & [O~{\scriptsize II}] \lam3727 \\
17    & 10 21 44.15 & $+$18 05 03.11 & 22.84 & 21.82 & 0.4763 & [O~{\scriptsize II}] \lam3727 \\
108   & 10 21 44.15 & $+$18 05 11.12 & 24.29 & 23.23 & 0.8320 & [O~{\scriptsize II}] \lam3727 \\
\,--  & 10 21 44.44 & $+$18 05 05.53 &\ldots &\ldots & 0.4763 & [O~{\scriptsize II}] \lam3727 \\
168   & 10 21 45.28 & $+$18 05 34.77 & 24.57 & 24.17 & 0.3603 & [O~{\scriptsize II}] \lam3727 \\
151   & 10 21 45.68 & $+$18 05 38.30 & 25.00 & 23.38 & 0.6028 & [O~{\scriptsize II}] \lam3727 \\
172   & 10 21 46.04 & $+$18 05 35.46 & 24.78 & 23.67 & 0.8149 & [O~{\scriptsize II}] \lam3727 \\
163   & 10 21 46.05 & $+$18 05 49.46 & 24.86 & 23.52 & 0.9692 & [O~{\scriptsize II}] \lam3727 \\
\,--  & 10 21 46.18 & $+$18 05 59.60 &\ldots &\ldots & 0.4137 & Ca~$H+K$ \\ 
\hline
\end{tabular}
\tablefoot{Identifiers and magnitudes are from \citet{m13}. MSB\,56
  was previously identified as a candidate C-AGB star by \citet{l16}.
  Typical uncertainties on the estimated redshifts are $\pm$0.0002,
  except for \#26 and 30, where the dispersion of the seven near-UV
  lines (Fe~{\scriptsize II} \lam\lam2343, 2374, 2382, 2586, 2599;
  Mg~{\scriptsize II} \lam\lam2796, 2803) was $\pm$0.0005.}}
\end{center}
\end{table*}

\section{Expected absolute magnitudes at low $Z$}\label{mags}

\begin{figure}
\begin{center}
\includegraphics[scale=0.5]{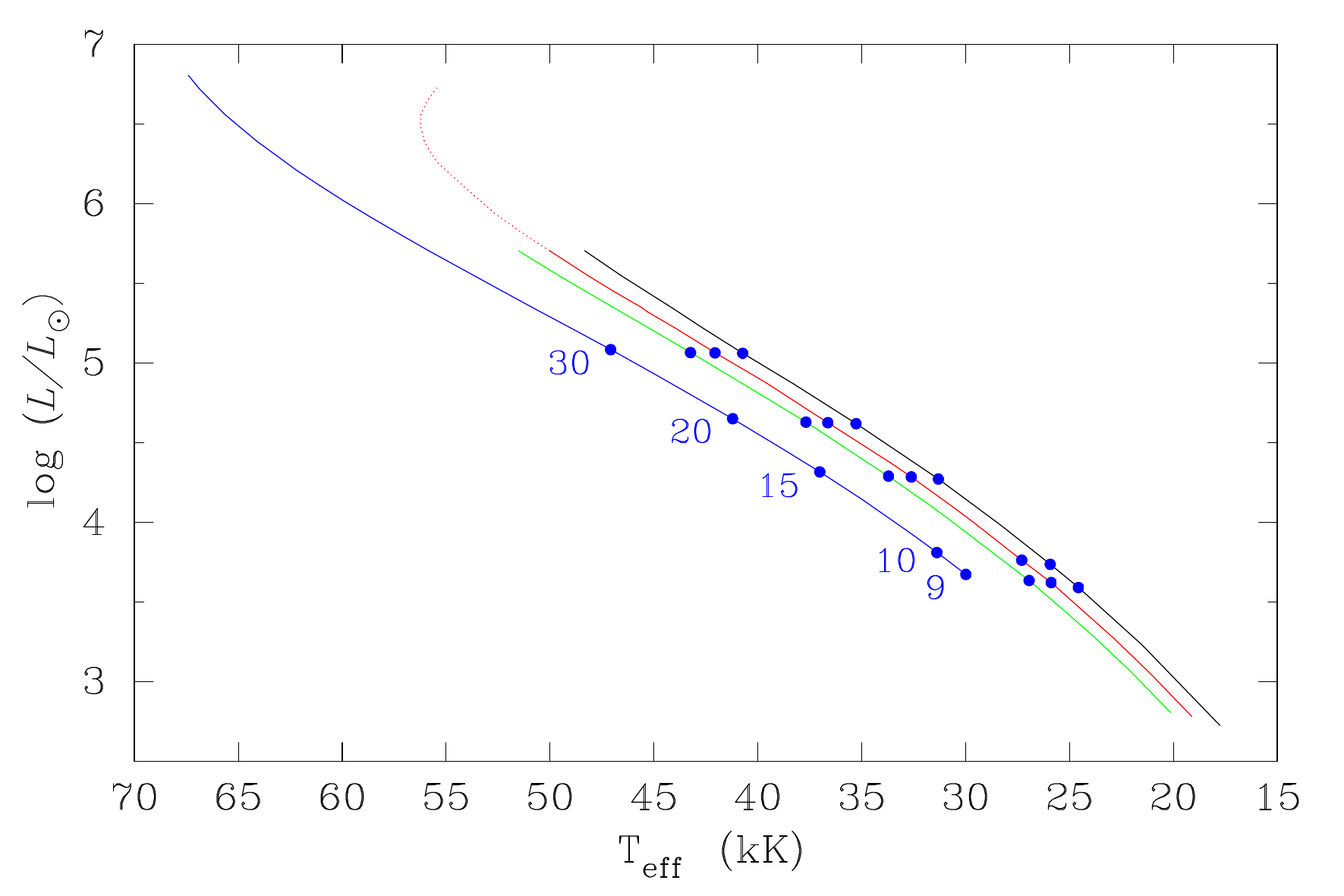}
\caption[]{Zero-age main sequence in the Hertzsprung--Russell diagram
  for (non-rotating) evolutionary models with $Z$\,$=$\,0.02$Z_\odot$
  \citep[blue,][]{s15}, and for the SMC (green), LMC (red), and Galaxy
  (black) from \citet{b11}; the LMC results are supplemented by
  higher-mass models (dotted red line) from \citet{k15}. Results for
  five initial masses (in units of M$_\odot$) are highlighted by the
  blue dots to illustrate the temperature dependence vs.
  metallicity (but with no 10\,M$_\odot$ model available for the SMC).}\label{tracks}
\end{center}
\end{figure}

We could potentially use the estimated absolute magnitude of LP\,26 to
(coarsely) constrain its spectral type, but we are unfortunately
limited to Galactic results \citep[e.g.][]{w72,w00,msh05}. To
investigate the potential impact of the very low metallicity in a
system like Leo~P on the expected stellar parameters we turned to
evolutionary models calculated by \citet{s15} for I\,Zw\,18 (with
$Z$\,$=$\,0.02\,$Z_{\odot}$). Fig.~\ref{tracks} shows the zero-age
main sequence (ZAMS) as traced by their non-rotating models compared
to results for the Galaxy, LMC and Small Magellanic Cloud (SMC) from
\citet{b11}; for completeness the LMC tracks are supplemented by
results for $M$\,$>$\,60\,$M_\odot$ from \citet{k15}. Also shown in
Fig.~\ref{tracks} are the ZAMS positions for models with initial
masses of 9, 10, 15, 20, and 30\,M$_\odot$, highlighting the
temperature dependence vs. metallicity for stars of a given mass (at
near constant luminosity, corresponding to smaller radii at lower
metallicities); similar differences are also seen in the rotating
models from the same grids. A shift in the same sense is also seen in
the `Geneva' evolutionary models, for example the comparison in Fig.~20 from
\citeauthor{s15} of the ZAMS for the SMC-like models of \citet{gee13}
and the near-zero metallicity models of \citet{mm02}.

As an example case, consider the ZAMS temperatures in
Fig.~\ref{tracks} for a 20\,M$_\odot$ star where T$_{\rm
  eff}$\,(MW)\,$=$\,35.3\,kK cf.  T$_{\rm
  eff}$\,(I\,Zw\,18)\,$=$\,41.2\,kK. The models at the ZAMS have
(near) constant luminosities, but such stars will have different
bolometric corrections (BCs) due to the increasing temperatures towards
lower metallicity. To investigate the scale of this effect we
highlight the BCs from the {\sc tlusty} OSTAR2002 grids \citep{lh03}.
A Galactic model for 35\,kK has BC\,$=$\,$-$3.24\,mag, cf.
BC\,$\sim$\,$-$3.8\,mag for the higher temperature at
$Z$\,$=$\,0.03\,$Z_{\odot}$ (interpolating between T$_{\rm
  eff}$\,$=$\,40 and 42.5\,kK).

There is a similarly large difference for the 9\,M$_\odot$ models,
ranging from 24.5\,kK in the Galaxy to 30\,kK for I\,Zw\,18. We again
turned to the {\sc tlusty} results to assess the potential impact of
this on the BC values, with BC\,$\sim$\,$-$2.95 for the hotter,
metal-poor star. The cooler, Galactic temperature is beyond the
OSTAR2002 grid, but an estimate of BC\,$\sim$\,$-$2.40 is available
from the companion BSTAR2006 grid \citep{lh07}. We require similar
calculations with the latest wind codes but these results illustrate
that, for a given mass, the visual absolute magnitudes of
high-mass stars at such low metallicities could be $\sim$0.5\,mag
fainter than we would otherwise expect (together with an increase in
the number of ionising photons).  Alternatively, this also implies
that fewer massive stars could potentially account for a given level
of ionisation in unresolved systems at very low-$Z$.

We are currently unable to link initial mass to spectral type as a
function of metallicity, but the differences above echo the effect
where metallicity is known to impact on the temperature scale of
OB-type stars of a given spectral type
\citep[e.g.][]{m05,m06,m07,t07}. At lower metallicities the cumulative
opacity of the metal lines is diminshed, resulting in less `back
warming' by trapped radiation and hotter models are required to
reproduce the observed ratios of spectral lines \citep[also see
discussion by][]{msh05}.

Taking this into consideration for the calibrations from
\citet{msh05}, the estimated absolute visual magnitude from
\citeauthor{m15} (M$_V$\,$=$\,$-$4.43) constrains the ionising source
in the H~\2 region in Leo~P to a mid O-type star. Given the prevalence
of binarity in massive stars, \citeauthor{m15} argued that this source
could perhaps be two O7 or O8 stars, with stellar masses of
$\sim$25\,M$_\odot$. Such a mass exceeds the maximum (of
2.5-3\,M$_\odot$) expected at the low star-formation rate of Leo~P
from the integrated galactic initial mass function (IGIMF) approach
\citep{pwk07}. Aside from the specific physical properties of LP\,26,
its spectroscopic confirmation as an O-type star supports the
conclusions of \citeauthor{m15} in this regard, i.e. that the upper
limit to the IMF -- even if sparsely sampled -- is not so
significantly influenced by the star-formation rate. More recently,
\citet{jhk18} have argued that the inferred star-formation rate in the
IGIMF theory for Leo~P would be significantly larger than the value
estimated from the current H$\alpha$ luminosity of the H~{\scriptsize
  II} region; this helps to reconcile the predictions with the
detection of the O-type star.

\section{Nebular emission}

\begin{figure}
\begin{center}
\includegraphics[width=9.75cm]{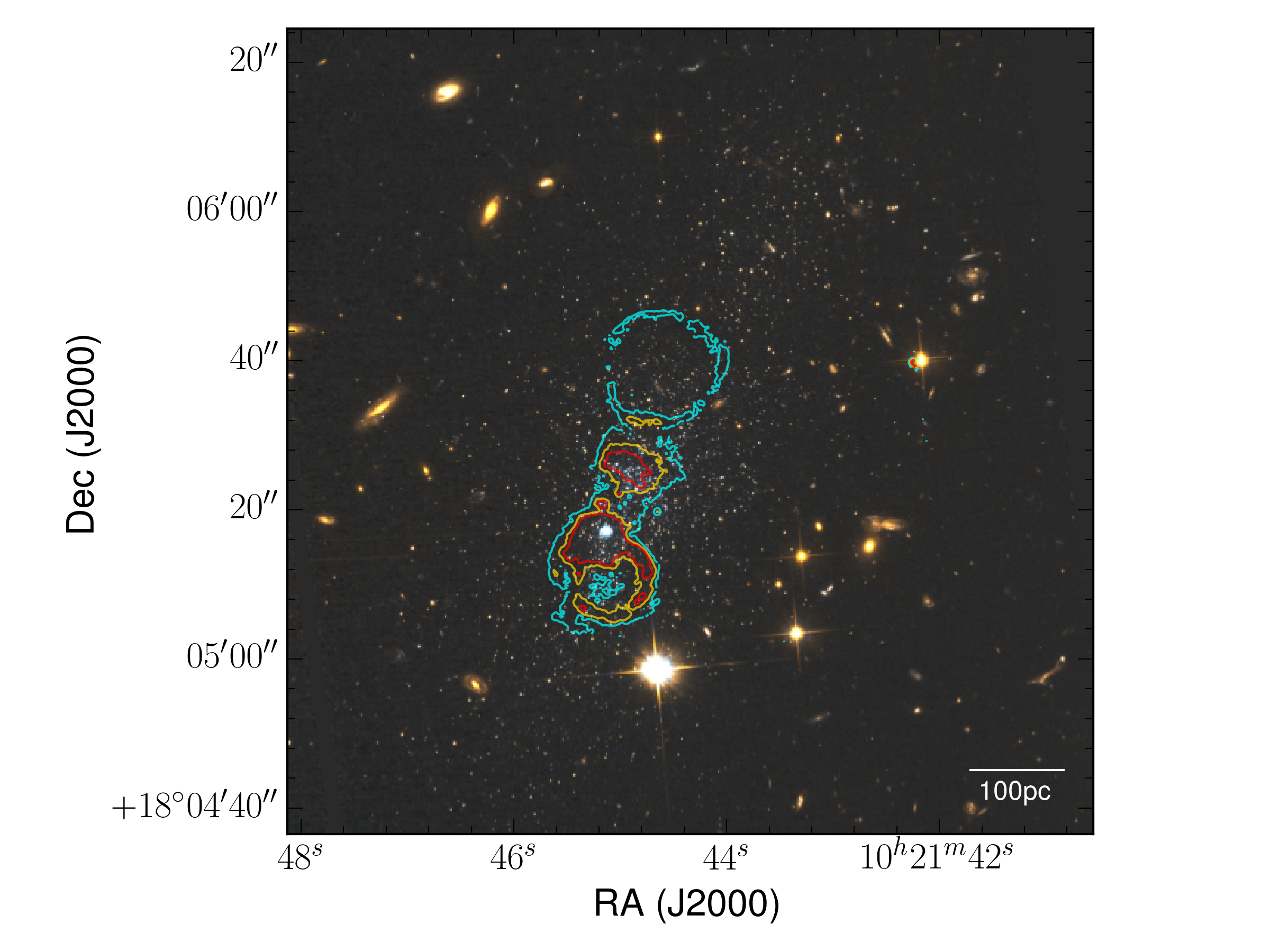}
\caption[]{H$\alpha$ emission in Leo~P overlaid on the combined
  $F475W+F814W$ {\em HST} image (cyan, yellow and red contours are
  H$\alpha$-emission levels of 5, 10, and
  15\,$\times$\,10$^{-20}$\,erg\,s$^{-1}$\,cm$^{-2}$,
  respectively). The known H~{\scriptsize II} region is the compact
  blue source in the main southern feature. The MUSE data have
  also revealed a region of H$\alpha$ emission just to the north, and
  two large extended shells to the south and (far) north (see also
  Fig.~\ref{emission_maps}).}\label{Ha_overlay}
\end{center}
\end{figure}

\begin{figure*}
\begin{center}
\includegraphics[width=18.5cm]{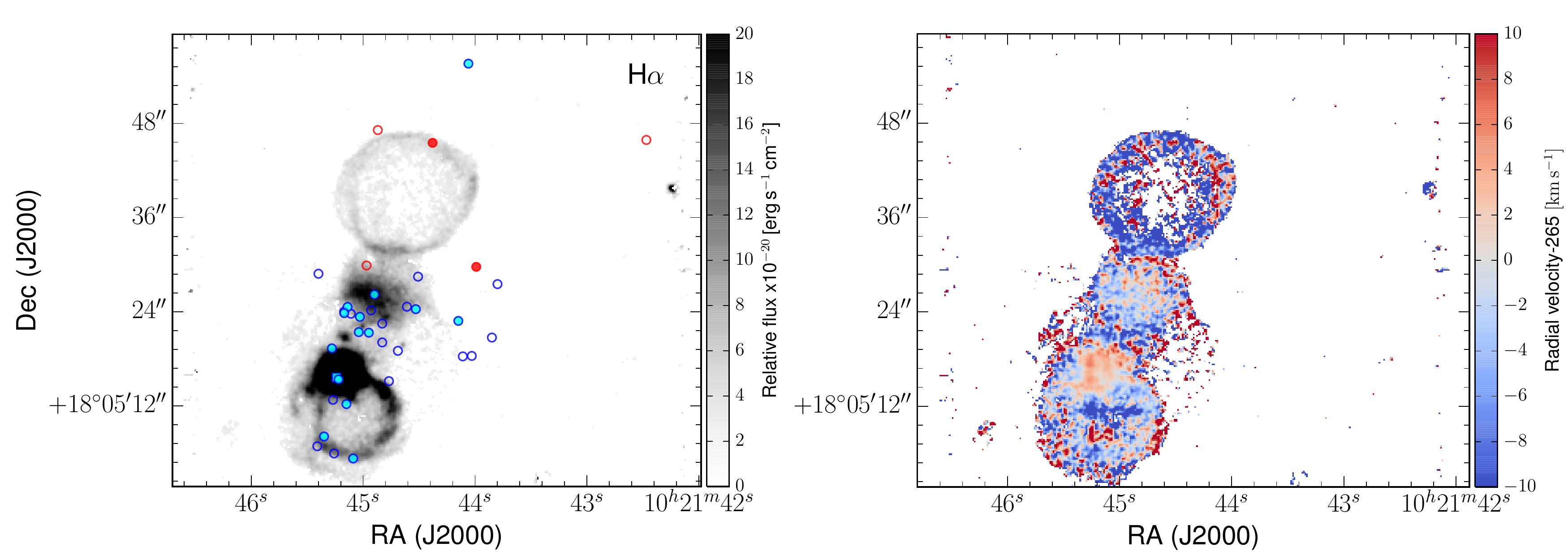}
\medskip
\includegraphics[width=18.5cm]{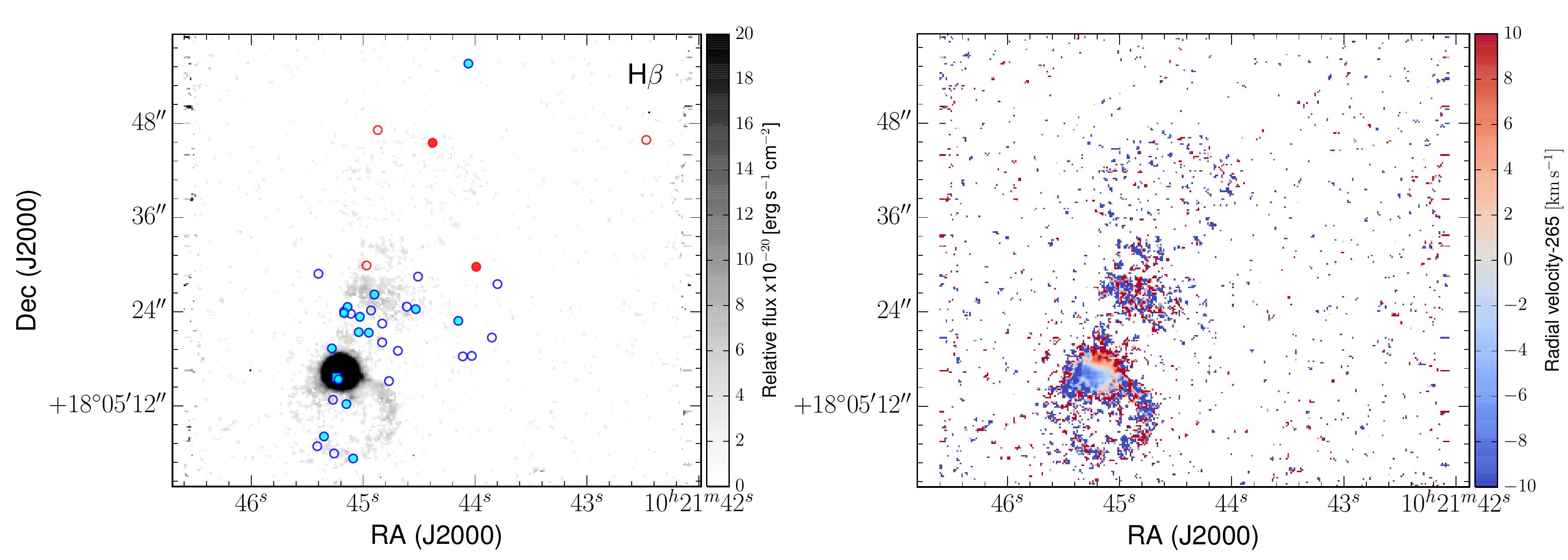}
\medskip
\includegraphics[width=18.5cm]{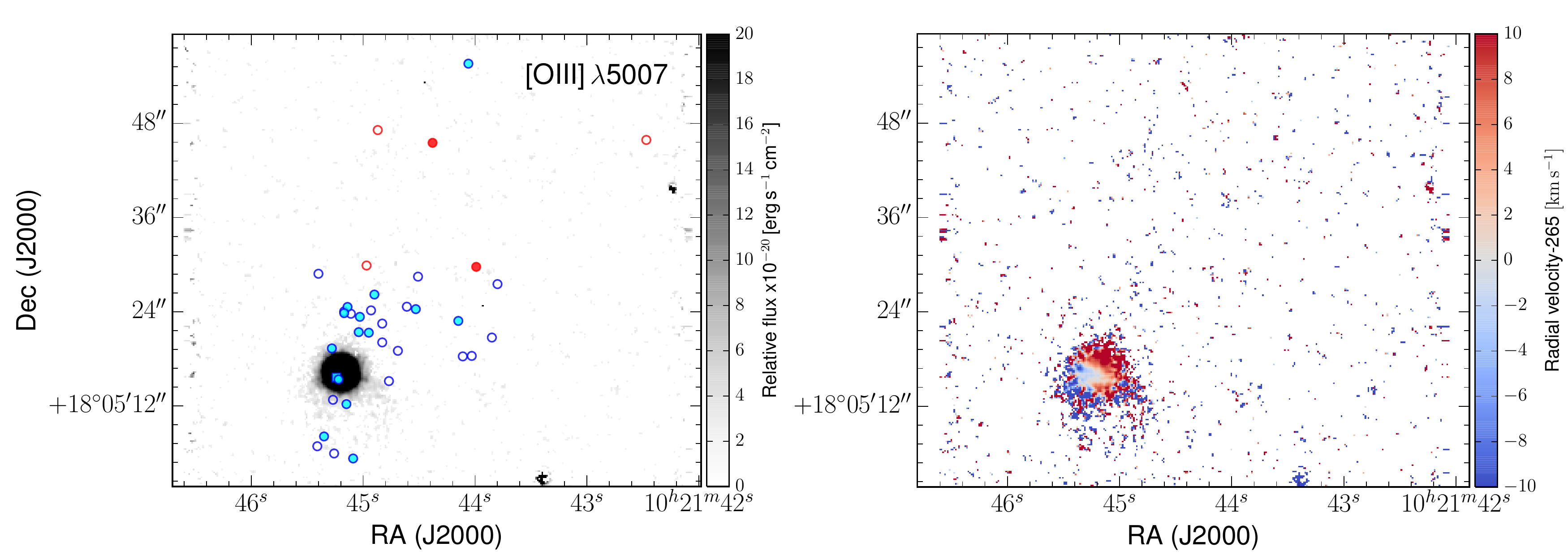}
\caption[]{Intensity and velocity maps for the ionised gas in Leo~P.
  The stars classified from the MUSE spectra are overplotted using the
  same symbols as in Fig.~\ref{spatial}, i.e.: massive stars (filled
  cyan circles), candidate massive stars (open blue circles), carbon
  stars (filled red circles), additional candidate AGB stars (open red
  circles). {\it Upper panels:} H$\alpha$ emission; {\it middle
    panels:} H$\beta$; {\it lower panels:} [O~{\scriptsize III}]
  \lam5007. {\it Left-hand panels:} intensity of the given emission
  line; {\it right-hand panels:} differential velocities from Gaussian fits to the
  observed lines.}\label{emission_maps}
\end{center}
\end{figure*}

Inspecting the MUSE data at the wavelength of H$\alpha$ emission we
discovered three new structures compared to the original discovery
images, suggesting more than one site of (relatively) recent star
formation, as shown in the contours in Fig.\ref{Ha_overlay} and the
upper panels of Fig.~\ref{emission_maps}. Part of the southern shell
is also traced by H$\beta$ emission (middle panels), while [O~\3]
\lam5007 emission is only seen in the vicinity of the H~\2 region
(lower panels). Intensity maps of [O~\3] \lam4959, [N~\2]
\lam\lam6584, 6548, and [S~\2] \lam\lam6717, 6731 were comparable to
that for [O~\3] \lam5007, in the sense that the large structures seen
in H$\alpha$ are absent (perhaps simply linked to the low
metallicity).

The northern H$\alpha$ shell has an apparent diameter of $\sim$15$''$.
Adopting a distance of 1.62\,Mpc \citep{m15}, this is equivalent to a
projected diameter of $\sim$120\,pc. The northern and southern shells
appear to be fairly well defined rings of emission, whereas a central
shell appears somewhat more diffuse. The H~\2 region is located on
northern edge of the southern ring.

The positions of our spectroscopically-confirmed stars are overlaid in
the left-hand panels of Fig.~\ref{emission_maps}. The absence of
massive stars in the northern shell suggests this is an older
formation (e.g. supernova remnant) rather than a more recent
wind-blown bubble. Only two of our sources lie within the southern
ring, including LP\,22, one of the candidate Be/Ae stars, which is
unlikely to be the driving source for such a ring, again suggesting it
as a supernova remnant. Indeed, given the large number of high-mass
x-ray binaries with Be-type components in the low-metallicity
environment of the SMC \citep[e.g.][]{c05}, we speculate that LP\,22
warrants further study to investigate for the presence of a degenerate
companion. Lastly, we also note the four stars along the southern edge
of the southern ring, suggestive of a connection in their formation.

Using a similar profile-fitting approach to that used by \citet{cce18}
for MUSE observations of 30~Doradus, velocity maps for each emission
line are shown in the centre panels of Fig.~\ref{emission_maps} (in
which the data are spatially smoothed with a Gaussian function with a
spatial FWHM of 0.6$''$ prior to the fits). We also investigated maps
of the standard deviation ($\sigma$) of the H$\alpha$ Gaussian fits to
investigate the velocity dispersion of the gas, but these do not add
much to the velocity maps, aside from hints of a slightly larger
dispersion in the northern shell.

The systemic velocity of the gas ($\sim$265\,\kms) is in good
agreement with the H~\1 and stellar velocities. There appears to be a
small velocity gradient of a few \kms\/ across the H~\2 region in the
H$\beta$ map, with a similar trend in [O~\3], but given the limitations
of the MUSE data, we defer detailed study of the kinematics to future
observations at higher spectral resolution \citep[e.g. with the Keck
Cosmic Web Imager,][]{kcwi}.

Prompted by the discovery of these spatially-distinct structures we
re-examined the {\em HST} photometry of all the stars in the four
regions (i.e. including those without spectroscopic confirmation from
MUSE). This led to relatively sparse samples and aside from
reinforcing the relative dearth of luminous, blue stars in the
northern and southern regions, we were unable to glean further
insights into the histories of these regions.

\section{Concluding remarks}

We have presented the first stellar spectroscopy in Leo~P, a low
luminosity, dwarf galaxy at a distance of 1.62\,$\pm$\,0.15\,Mpc. Our
findings from the MUSE observations include:
\begin{itemize}
\item{Spectroscopic confirmation of an O-type star (LP\,26) in the
    H~\2 region via observations of He~\2 absorption. From
    consideration of its absolute magnitude (and assuming the
    published distance), this is probably a mid O-type star
    (Fig.~\ref{HIIregion}). }
\item{Fourteen sources with H$\beta$ absorption and rising blue flux
    distributions that confirms them as hot stars. Given the faintness of
    these targets (22\,$<$\,$V$\,$<$\,25) we were unable to comment
    more on their spectral classifications, but from their locations in
    the CMD we suggest these are B-type (or late O-type) objects
    (Fig.~\ref{blue}). We tentatively classify a further 17 sources as
    candidate hot stars, again via detection of H$\beta$ absorption.}
\item{Confirmation of two candidate AGB stars from \citet{l16} as
    carbon stars, and confirmation of two further candidates as
    luminous cool (presumably AGB) stars via detection of CaT
    absorption in their spectra (Figs~\ref{agb_lee} and \ref{CaT}).}
\item{Confirmation of CaT absorption in the RGB population, via
    co-adding the spectra of the brightest members (Fig.~\ref{CaT}).}
\item{Two 100\,pc-scale ring structures that are traced by
    H$\alpha$-emission from the gas, with the H~\2 region located on
    the northern edge of the southern ring (Fig.~\ref{emission_maps}).}
\end{itemize}

In addition, to investigate the expected temperatures and magnitudes
of massive stars in Leo~P, we employed evolutionary tracks from
\citet{s15} for single stars at the (near identical) metallicity of
the I\,Zw\,18 galaxy. The shift of the ZAMS to higher temperatures with
decreasing metallicity has long been known, but we argue that the
significance of this effect at very low metallicites could impact on
the apparent magnitudes of massive stars. For a given stellar
luminosity, hotter temperatures give larger bolometric corrections,
which would mean that stars of a comparable mass in Leo~P or I\,Zw\,18
will have fainter absolute visual magnitudes than in the Galaxy (by
$\sim$0.5\,mag in the example cases of the 9 and 20\,M$_\odot$ models
considered here).

While giving us our first tantalysing view of high-mass stars in a
very metal-poor environment, the upper mass function is so sparsely
populated that we were not able to test the predictions of \citet{s15}
regarding core-hydrogen-burning supergiants and blueward evolution of
stars from the ZAMS (caused by chemically-homogeneous evolution).
Nonetheless, the MUSE observations have given us a first census of the
high-mass population of Leo~P, providing confirmed targets for
long-exposure spectroscopy to obtain higher S/N and better spectral
resolution. In particular, we highlight future observations of the
O-type star (LP\,26) to derive its physical properties and a more
robust estimate of its mass (cf. predictions from the IGIMF).
Moreover, to interpret such observations we will also require
synthetic spectra at this low metallicity from the latest
model-atmosphere codes.

The MUSE data have also provided a first look at the cool, evolved
population of Leo~P. Quantitative analysis to determine their physical
parameters (in part to calibrate low-$Z$ evolutionary models) and
detailed kinematic analysis to investigate the dynamical properties of
the cool population will again require ambitious spectroscopic
follow-up.

Looking further into the future, the Extremely Large Telescope (ELT)
will have the combination of both angular resolution and sensitivity
to probe the evolved population of Leo~P in much greater depth, for
example with the first-light HARMONI visible and near-IR, integral-field
spectrograph \citep{harmoni}. Ultimately we also want ultraviolet
spectroscopy of the population of massive stars in Leo~P, to
investigate their wind properties as well as their physical parameters
in this important low-metallicity regime. This is unrealistic at
present with {\em HST}, but would be well within the grasp of the
proposed Large Ultraviolet Optical Infrared Surveyor (LUVOIR) concept
currently under study \citep[e.g. see Section 4.3 from][]{luvoir}.

\begin{acknowledgements}
  Based on observations at the European Southern Observatory Very
  Large Telescope in programme 094.D-0346. We thank the referee for
  their constructive and helpful suggestions, and are also grateful to
  Olivia Jones, Anna McLeod and Rub\'{e}n S\'{a}nchez-Janssen for
  useful discussions in the course of this work. DRW is supported by a
  fellowship from the Alfred P. Sloan Foundation, and acknowledges
  support from the Alexander von Humboldt Foundation. This research
  made use of Astropy, a community-developed core Python package for
  Astronomy \citep{astropy}, and APLpy, an open-source plotting
  package for Python \citep{rb12}.
\end{acknowledgements}

\bibliographystyle{aa}
\tiny
\bibliography{34145}

\end{document}